# Content-based data leakage detection using extended fingerprinting

Yuri Shapira   Bracha Shapira   Asaf Shabtai
Dept. of Information Systems Engineering
Ben-Gurion University of the Negev
Israel
*{shapira, bshapira, shabtaia}*@bgu.ac.il

## ABSTRACT
Protecting sensitive information from unauthorized disclosure is a major concern of every organization. As an organization's employees need to access such information in order to carry out their daily work, data leakage detection is both an essential and challenging task. Whether caused by malicious intent or an inadvertent mistake, data loss can result in significant damage to the organization. Fingerprinting is a content-based method used for detecting data leakage. In fingerprinting, signatures of known confidential content are extracted and matched with outgoing content in order to detect leakage of sensitive content. Existing fingerprinting methods, however, suffer from two major limitations. First, fingerprinting can be bypassed by rephrasing (or minor modification) of the confidential content, and second, usually the whole content of document is fingerprinted (including non-confidential parts), resulting in false alarms. In this paper we propose an extension to the fingerprinting approach that is based on sorted $k$-skip-$n$-grams. The proposed method is able to produce a fingerprint of the core confidential content which ignores non-relevant (non-confidential) sections. In addition, the proposed fingerprint method is more robust to rephrasing and can also be used to detect a previously unseen confidential document and therefore provide better detection of intentional leakage incidents.

## Categories and Subject Descriptors
C.2.0 [**COMPUTER-COMMUNICATION NETWORKS**]: Security and protection

## General Terms
Security

## 1. INTRODUCTION
One of the most dangerous threats that organizations are facing today is leakage of confidential data [20]. Data leakage is defined as an unauthorized transfer of sensitive data from within an organization to an unauthorized external destination or recipient [48]. The financial damage caused by the leakage of sensitive data can be significant. According to the Ponemon Institute 2010 annual study [33], the average cost of a compromised customer record ranges from 4 to 156 USD. Customer private information that is leaked out can result in loss of reputation, customer abandonment and even fines, settlements or customer compensation fees. Two recent Sony PlayStation Network data breaches are excellent examples of this scenario. The first breach compromised 77 million user accounts [4] and the second breach compromised another 25 million [3]. Such an accident apparently caused the customers to turn to the competitors [15] in addition to Sony's lawsuit [17].

Data leakage incidents can be classified as sourcing from outside or inside the organization. Furthermore, data leakage by insiders can be categorized as intentional or unintentional. The most challenging task is to identify *intentional* data leakage by *insiders* who may hold legitimate access rights to the confidential data. Two examples of such a case are an employee of an organization sending an email to a friend describing a new project she is working on or an accountant sending still unpublished revenue reports using some instant messaging application. According to the study [2] conducted by the Verizon RISK Team in cooperation with the United States secret services, 48% of data breaches in 2010 were caused by insiders.

In modern organizations there are generally several types of digital data that can potentially be leaked: structured data (tables, lists, etc.), textual data (reports, contracts, emails, etc.), technical drawings, and multimedia (images, audio and video). However, the methods to detect and prevent leakage of each type differ significantly. The method proposed in this paper is aimed at detecting leakage of textual data.

Usually, data leakage detection (DLD) solutions identify confidential data using the following three approaches:
- *Context based* – inspects contextual information extracted from the monitored data (source, destination, file type, time stamp, etc.) with an intention to find anomalies or to match to one of the predefined security policy rules; e.g., sending source code file outside the organization is prohibited.
- *Content tagging* – assigns a tag to a file containing confidential data and a policy is enforced based on the assigned tag. Content will remain tagged even when processed by other application (e.g., compressed).
- *Content based* – detects a leakage by analyzing the content of file.

In this paper we describe a content-based method that can be used for detecting data leakage sourcing from inside the organization and mitigate the intentional data leakage scenario specifically.

The most well-known approach for content-based detection is *fingerprinting*. In fingerprinting, a known confidential document (content) is converted into a set of hash values. The inspected document (content) is also fingerprinted and compared with these hash values. Even though fingerprinting is a very effective data leakage detection method, it has the following limitations [40].
- Since each fingerprint is computed from a sequence of words (or letters), a change in one character of the fingerprinted text will result a different hash value and can therefore be bypassed by rephrasing (or minor modification) of the confidential content by replacing all characters 'O' to digits '0', for example. In order to overcome this problem, existing DLD solutions create fingerprints of each confidential file with each modification or combination of few modifications, but this solution is very limited.
- Usually, the whole content of a document fingerprinted, including standard content (disclaimers, standard form, common phrases, etc.) This may result in false alarms, when the standard content appear in outgoing document. An example can be seen in Figure 1. Current DLD systems handle this problem by manually managing a whitelist of such content, but this requires a lot of human resources.

Fingerprinting is extensively used in plagiarism detection [38], [24], [6], finding near-duplicate files [30], [26], [29] and authorship detection [1]. However, to the best of our knowledge, there are no academic studies related to the application of fingerprinting to data leakage detection. Data leakage detection is closely related to the above mentioned problems and therefore fingerprinting methods are also appropriate for detecting leakage of confidential data. However, due to the following special properties of the DLD domain fingerprinting should be adjusted:
- In the DLD domain two sets of document can be used - confidential and non-confidential. When existing fingerprinting methods are applied to data leakage detection domain the non-confidential document set is overlooked. Nevertheless, considering both sets of documents may add necessary information for the process of distinguishing between confidential and non-confidential content, thus reducing false alarms count.
- The documents in the confidential document set usually relate to a limited number of pre-specified topics, providing the ability to detect previously unseen confidential content by classifying it to one of those topics.
- To assure the feasibility of the DLD method, it should be sufficiently efficient to allow examining documents at real-time, as opposed to plagiarism detection, for example, that can be performed offline.

We propose a straightforward extension to the fingerprinting approach that enables extraction of fingerprints from the core confidential content while ignoring non-relevant (non-confidential) parts of a document. In addition, the proposed fingerprint method is more robust to rephrasing and can even be used to detect a previously unseen confidential document (which is actually not the main purpose of the proposed method), and therefore provide better detection of intentional leakage incidents. The proposed method is simple and easy to apply and yet outperforms the state-of-the-art fingerprint methods.

It is common practice to use *n*-grams as features for representing documents [8], [24], [38]. We use *k*-skip-*n*-gram which is an *n*-gram with up to *k* "skips" between its elements (referred to as skip-grams) [19]. Skip-grams were previously used in domains that require robustness to noise in the data, such as speech recognition [32], language modeling [19], plagiarism detection [21], and automatic evaluation of text summaries [44]. Furthermore, we sort the words in each skip-gram in alphabetic order.

These changes increase the robustness of fingerprinting to rephrasing. When some of the words are replaced by synonyms and/or the order of words in the sentence is changed, the confidential content can still be detected using sorted skip-grams. To the best of our knowledge, the proposed method is the first to use skip-grams for text fingerprinting.

In order to reduce the amount of non-confidential content within a document fingerprint, only skip-grams that appear in less than *m* non-confidential documents are considered as features for a document's fingerprint.

Therefore, the main contribution of this paper is the presentation of an extended version of full fingerprinting [24] that is more robust to the rephrasing of confidential content and can filter non-relevant (non-confidential) parts of a document. We present an evaluation of the proposed method and illustrate its superiority over the full fingerprinting approach while maintaining acceptable execution time.

The rest of the paper is organized as follows. In section 2 we introduce basic definitions. In section 3 we survey popular content-based data leakage detection approaches. In section 4 we describe in detail a new fingerprinting method for data leakage detection. In section 5 we present preliminary evaluation results of the proposed method and its comparison with the well-known full fingerprinting method [24]. Section 6 concludes the paper.

| Fingerprinted confidential document | Outgoing document |
|---|---|
| Irene, | Hi Jack, |
| What is the status of the pilot timelines? Last week you mentioned that you were waiting for Henry to send you the development timeline for the pilot, and that you were working on communication and planning documents (including timelines) for the pilot. Your assistance in expediting this information would be appreciated. | Do you want to play poker tomorrow night at my place? **John Smith** T: +111.51.23989132 F: +111.51.23989133 E: john@company.com http://www.company.com |
| Thank you as always for your cooperation. | |
| **John Smith** | |
| T: +111.51.23989132 | |
| F: +111.51.23989133 | |
| E: john@company.com | |
| http://www.company.com | |

**Figure 1. Standard content illustration.**

## 2. DEFINITIONS
We introduce some definitions for the types of data leakage challenges that content-based DLD method should be able to handle.

Given a document $d$ with content $T_d$ and a set of confidential documents $S$:
- *Full duplicate detection* ($\exists\, s \in S: T_d = T_s$) – exact copy of confidential document.
- *Near duplicate detection* ($\exists\, s \in S: T_d \approx T_s$) – confidential document with a small number of differences.

- *Partial duplicate detection* ($\exists\, s \in S \wedge \exists\, t_d \subseteq T_d \wedge \exists\, t_s \subseteq T_s: t_d = t_s$) - small portions of confidential text are embedded in a larger, non-confidential text.
- *Rephrased text segment detection* ($\exists\, s \in S \wedge \exists\, t_d \subseteq T_d \wedge \exists\, t_s \subseteq T_s: t_d \approx t_s$) - small portions of modified (rephrased) confidential text are embedded in a larger, non-confidential text Detection.
- *Previously unseen text detection* ($topic(d) \in topics(S)$) – previously unseen confidential document.
- *Standard content handling* ($\exists\, s \in S \wedge \exists\, t_s \subseteq T_s: t_s$ is not confidential) - ignore non-confidential content (disclaimers, standard forms, common phrases, etc.), in spite of its presence within confidential documents.

## 3. CONTENT-BASED DATA LEAKAGE DETECTION

There are a few popular content-based approaches that are used for detecting data leakage in outgoing information [40], [46]:
- Global filters - concerning the whole file:
  o File-based binary signature – hash value of the whole file. Can detect only an exact copy of a confidential file.
  o Text-based binary signature – hash value of textual content of file. Offers more robustness compared to previous method. Can detect converted files e.g., txt to doc. Ignores text metadata like font, color, etc.
- Tokens - concerning special keywords or patterns:
  o Keywords filter – used to build a policy based on keywords. For example, block files that mention "Project X".
  o Pattern recognition – can block documents containing a match to a credit card number, phone number, etc.
- Machine learning – can detect previously unseen confidential documents. Machine learning methods classify the documents according to their similarity to confidential or non-confidential documents.
- Textual fingerprint – can detect full, near, and partial duplicates of confidential documents.

Table 1 presents a comparison of the detection abilities of each method.

Global filters suffer from very low robustness e.g., even if a single character is replaced in a file, it cannot be detected. Token-based filters require a very accurate selection of keywords and patterns, which is not a realistic option in most cases. Machine learning (ML) methods require a lot of documents for training and suffer from high false positive rates. Additionally, ML methods cannot detect a partial duplicate of a confidential document since most of the document content is non-confidential and ML is based on statistics. Therefore, for content-based data leakage detection, the most natural choice is textual fingerprint. It is an efficient and effective approach and is commonly used by leading commercial DLD products; although, it also has some limitations as explained later in the section.

A fingerprint of a document is a set of hash values of its features. In order to check whether document $d$ is similar to one of the documents in a reference set $R$ (when fingerprint is applied to data leakage detection, the set $R$ consists of the confidential documents of the organization) indexing and detection phases are required. A pre-processing (indexing) phase is applied where each of the documents in $R$ is fingerprinted. The fingerprints are stored in a special database (Figure 2). Then, during the detection phase, a fingerprint of the examined document $d$ is extracted and is compared with fingerprints in the database. A list of documents that contain each of the hashes that are included in the fingerprint of document $d$ is efficiently retrieved (using inverted index) from the database. The documents that share a number (above some predefined threshold) of hashes with $d$ are considered as similar. Therefore, there is no need to make a pairwise comparison of each document in $R$ with $d$ and the process time is linear to the length of $d$. This property makes fingerprinting highly scalable. As a case in point, Google's crawler employs their fingerprinting implementation to detect near duplicate web pages, while the reference set is the part of the Internet that Google is indexing [31]. Thus, it is naturally appropriate for real time environment.

**Table 1. Comparison of content-based methods appropriate for data leakage detection.**

|  | Global Filters | Token-Based | Machine Learning | Fingerprinting | | | | |
|---|---|---|---|---|---|---|---|---|
|  |  |  |  | Classic Fingerprinting | LSH Based | Collection Statistic Based | Anchor Based | Proposed Method |
| **Full Duplicate** | ✓ | ⫟ | ✓ | ✓ | ✓ | ✓ | ✓ | ✓ |
| **Near Duplicate** | ✗ | ⫟ | ✓ | ✓ | ✓ | ✓ | ✓ | ✓ |
| **Partial Duplicate** | ✗ | ⫟ | ✗ | ✓ | ✗ | ⫟ | ✓ | ✓ |
| **Rephrased Text Segment** | ✗ | ⫟ | ✗ | ⫟ | ✗ | ⫟ | ⫟ | ✓ |
| **Previously Unseen Text** | ✗ | ⫟ | ✓ | ✗ | ✗ | ✗ | ✗ | ⫟ |
| **Standard Content Handling** | ✓ | ✓ | ⫟ | ✗ | ⫟ | ✗ | ⫟ | ✓ |

✓ - effective solution ⫟ - partial solution ✗ - not appropriate

Most fingerprinting techniques select only a small subset of the fingerprinted document's terms in order to minimize the fingerprints database size without significantly affecting the accuracy. Surprisingly, it turns out that in some cases, considering many (or even all) terms of a document for its fingerprint does not improve the accuracy, possibly due to noise (stopwords, tags, etc.) [24]. Therefore, fingerprinted terms should be carefully selected for an optimal representation of a document. Previous studies e.g., [8], [38], and [22], focus on the strategy for selecting terms for the fingerprinting process. The following is a description of existing fingerprinting methods and their advantages and disadvantages.

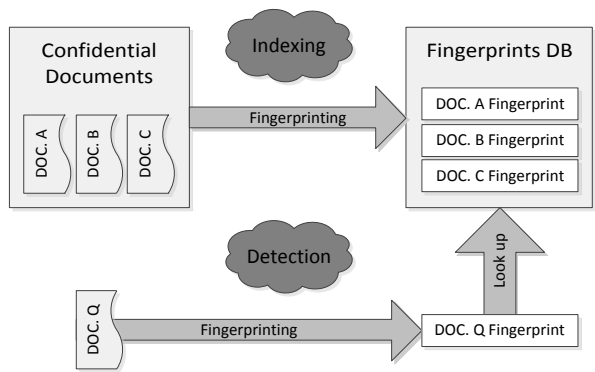

**Figure 2. Finding similar documents with fingerprinting.**

**Classic fingerprinting** [30], [8] [10], [38], [42]

Literally, the term *"fingerprinting"*, in most cases, refers to a schema with the following four processing steps:

- *Feature extraction* – input document is split into tokens (characters, words, sentences, etc.) Then, text pre-processing techniques (stemming, stopwords removal, etc.) are applied. The influence of different techniques when fingerprinting is applied to plagiarism detection discussed in [24] for English and in [12] for Czech.
- *Feature separation* – overlapping *n*-grams extracted from the sequence of tokens.
- *Feature hashing* - *n*-grams are hashed using MD5 [37], SHA [43], Rabin [36] or another hash function.
- *Hash selection* – the goal of this step is to reduce the document fingerprint size. A few selection strategies were proposed, the best known are 0 mod n [8] and winnowing [38]. Fingerprinting without this step is known as Full Fingerprinting [24].

**LSH-based fingerprinting** [9], [7], [13], [31], [23], [5]

These techniques are based on Locality Sensitive Hashing ([25], [18]) which produces similar hash values when applied on similar objects so that the similarity between objects can be estimated by the similarity between hash values. These methods produce a very compact representation of documents, but were developed with the intention of detecting near-duplicate documents and can hardly deal with detection of a small portion of shared content between documents.

**Collection statistic-based fingerprinting** [41], [22], [14], [29]

These techniques select terms using collection statistics. The main assumption is that terms that are rare in the whole collection and common in the current document very well describe the main idea of the document. This assumption strongly reminds us of *tf-idf* [27] measure from information retrieval.

**Anchor based fingerprinting** [30], [45], [49]

Select sequences of terms that start from one of a predefined set anchor words. These anchor words should be carefully selected to be common in the core content of a document while at the same time rare in document framing (ads, disclaimer, etc.), therefore it is better to select them for every specific domain.

There are also other fingerprinting approaches, such as using the *fast Fourier transform* (FFT) to create fingerprints robust to small changes [39], adding small *bloom filter* to documents to allow fast similarity check [26], and a method for detecting documents derived from templates [28]. However, all of these methods are not appropriate for detecting rephrased confidential content.

## 4. THE PROPOSED METHOD

### 4.1 Background

*N*-grams matching is a well-known method that was applied in a diverse of fields such as text categorization [11] and music information retrieval [16]. *N*-gram is a slice of *n* elements extracted from a longer sequence of elements.

Most of fingerprinting methods represent documents as a set of hash values of its *n*-grams of terms [24] where *n* refers to the number of terms in the slice. For example, the 3-grams of the news title "*Chinese activist appeals to the United States for help*":

> ***Chinese activist appeals*** *to the United States for help*
> *Chinese **activist appeals to** the United States for help*
> *Chinese activist **appeals to the** United States for help*
> *...*

*k*-skip-*n*-gram is an *n*-gram that allows up to *k* "skips" between its elements; e.g., for the above title:

- 0-skip-3-grams:

  > ***Chinese activist appeals*** *to the United States for help*
  > *Chinese **activist appeals to** the United States for help*
  > *Chinese activist **appeals to the** United States for help*
  > *...*

- 1-skip-3-grams: all 0-skip-3-grams and:

  > ***Chinese*** *activist **appeals to** the United States for help*
  > ***Chinese activist*** *appeals **to** the United States for help*
  > *Chinese **activist** appeals **to the** United States for help*
  > *...*

- 2-skip-3-grams: all 1-skip-3-grams and:

  > ***Chinese*** *activist appeals **to the** United States for help*
  > ***Chinese activist*** *appeals to **the** United States for help*
  > ***Chinese*** *activist **appeals** to **the** United States for help*
  > *...*

*Sorted k*-skip-*n*-gram is a *k*-skip-*n*-gram which elements are sorted in alphabetic order.

The skip-grams add contextual information that cannot be achieved using standard *n*-grams since the context of term can be better represented by considering not only the adjacent terms. The exact number *J* of *k*-skip-*n*-grams extracted from a sequence of terms of length *L* can be calculated using the following formula:

$$J = \sum_{i=0}^{k}(L-n-i+1)\frac{(n-2+i)!}{(n-2)!\,i!}, for\ L > k+2$$

For $n,k \ll L$ the reduced number of skip-grams extracted from few last terms in the sequence have low effect on the overall number of skip-grams, so $J$ can be approximated:

$$J \approx L \cdot \frac{(k+n-1)!}{k!\,(n-1)!}$$

## 4.2 An Overview of the Proposed Method

The presented fingerprinting based DLD has two main processing phases, as can be seen in Figure 3:
- *Indexing phase* – fingerprints extracted from confidential documents and stored into the database. As opposed to existing methods, the proposed fingerprinting process considers non-confidential documents as well, as will be explained later.
- *Detection phase* – fingerprint from outgoing documents extracted and compared with confidential fingerprints in the database.

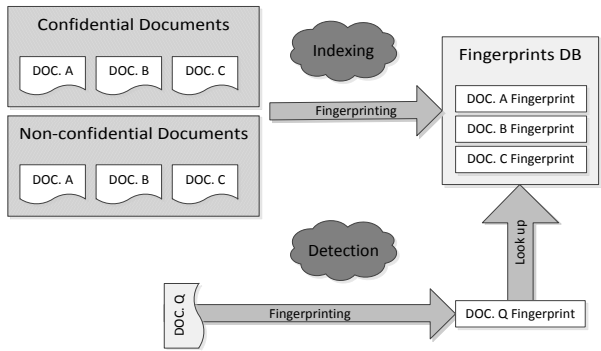

**Figure 3. Proposed method's phases.**

In modern organization, the document sets that represent confidential and non-confidential information may change each day. For example, highly sensitive information about new product becomes publicity available after its launch. Furthermore, large organization produces enormous amount of sensitive reports/emails/presentations which leakage attempt should be detected right after creation. To support the dynamic nature of the dataset the applied detection method should enable real-time performance of indexing new documents and changing the status of confidential documents to non-confidential. In the proposed method these operations are linear to the file's size, since they require only an addition or deletion from a database.

## 4.3 Indexing Phase

The input of this step consists of two document sets:
- **Confidential documents** – the documents which contain confidential content (while still may have portions of non-confidential content).
- **Non-confidential documents** – the documents which do not contain any confidential content.

Generally, we need the organization to correctly tag the input documents, but in actuality, the proposed method allows for a small percent of noise (errors) in tagging, making it appropriate for real life noisy input.

Both confidential and non-confidential documents undergo fingerprinting using the classic fingerprinting schema with adjustments to the feature separation and hash selection steps. The following is a description of the fingerprint schema we used.
- *Feature extraction* – input documents are split into words, while words are defined as sequences of characters or digits. Then, optionally, words are stemmed using Porter stemmer [34] and stopwords removed. According to our preliminary evaluation, stopwords removal usually degrades detection accuracy, confirming the results of a recent work on Plagiarism Detection [42]. Therefore, stopwords should be removed only when the main intend is previously unseen text detection.
- *Feature separation* – This step is adjusted. We applied sorted $k$-skip-$n$-grams extraction on the word sequence from the previous step. The $k$-skip-$n$-grams are non-continuous version of $n$-grams, where up to $k$ skips are allowed. Thus, 0-skip-$n$-grams are equal to $n$-grams. Then, the words in the skip-grams are sorted in alphabetic order.
- *Feature hashing* – $n$-grams are hashed using .Net string hash function[1]. The preliminary evaluation did not show a significant effect of the chosen hash function on the performance of the developed method.
- *Hash selection* – This step is applied to confidential documents only. This step is modified so that only hashes which appear in less than $m$ non-confidential documents are considered as a document's fingerprint. The idea is to make a fingerprint of a core confidential content of a confidential document, ignoring common phrases, disclaimers, standard forms, etc. According to our preliminary evaluation, m should be set to 1, i.e., even if a skip-gram appears only in a single non-confidential document, it is worthwhile to filter it out of the fingerprints of confidential documents. However, in order to make a method robust to errors in the tagging of input documents, $m$ should be set to a higher value.

The fingerprints of confidential and non-confidential documents are stored into database. At the detection phase only fingerprints of confidential documents are used so when the indexing and detection are executed on different computers, only the fingerprints of confidential documents should be synchronized.

The demonstration of the indexing phase using the new method and with full fingerprinting [24] can be seen in Figure 4. In this example, both confidential and non-confidential documents talk about Barack Obama's invitations to the White House, however only the confidential document talks about Barack Obama's invitation of Netanyahu. Therefore, it can be concluded that information about Barack Obama's invitation is not necessarily confidential and should not cause a false alarm in the detection phase. As can be seen, the proposed method ignores "*barack invit obama*" and "*hous visit white*", implementing this intuition. As more non-confidential documents are provided, the new method will better distinguish between confidential and non-confidential content.

---

[1] http://msdn.microsoft.com/en-us/library/system.string.gethashcode.aspx

| | Full Fingerprinting | New Method |
|---|---|---|
| Settings | *n*=3 | *n*=3, *k*=1 |
| Input | Confidential document: "Barack Obama invites Netanyahu for White House visit." | Confidential document: "Barack Obama invites Netanyahu for White House visit." Non-Confidential document: "President Barack Obama has invited the Dalai Lama to visit the White House" |
| Fingerprint* | barack obama invit<br>obama invit netanyahu<br>invit netanyahu white<br>netanyahu white hous<br>white hous visit | barack netanyahu obama<br>invit netanyahu obama<br>invit obama white<br>netanyahu obama white<br>invit netanyahu white<br>hous invit netanyahu<br>hous invit white<br>hous netanyahu white<br>netanyahu visit white<br>hous netanyahu visit<br>Ignored skip-grams:<br>barack invit obama<br>hous visit white |

*Actually, the hashes of the presented n-grams/skip-grams are stored

**Figure 4. Indexing phase example. Comparison of full fingerprinting with the new method.**

| | Full Fingerprinting | New Method |
|---|---|---|
| Settings | *n*=3 | *n*=3, *k*=1 |
| Input | "Barack Obama has issued an invitation to Israeli Premier Benjamin Netanyahu to visit the White House." | |
| Fingerprint* | barack obama issu<br>obama issu invit<br>issu invit isra<br>invit isra premier<br>isra premier benjamin<br>premier benjamin netanyahu<br>benjamin netanyahu visit<br>netanyahu visit white<br>visit white hous | barack invit obama<br>barack invit issu<br>invit issu obama<br>isra issu obama<br>invit isra obama<br>invit isra issu<br>invit issu premier<br>isra issu premier<br>... |
| Matches | | netanyahu visit white<br>hous netanyahu visit<br>hous netanyahu white |

*Actually, the hashes of presented n-grams/skip-grams are compared

**Figure 5. Detection phase example. Comparison of full fingerprinting with the new method.**

## 4.4 Detection Phase

The input of this processing step is a document *d* and the output is the "confidentiality score" of *d*. A document with a "confidentiality score" above some threshold is considered confidential and is detected as leakage.

The input document *d* is fingerprinted using the new fingerprinting method (described in section 4.2) resulting in a list of hashes representing the document. Then, a list of documents that contain at least one of these hashes is retrieved from the fingerprints database. The confidentiality score of document *d* is set to the maximal number of hashes detected in any of the documents in the list.

The proposed method is more robust to rephrasing than the traditional methods, since the use of sorted skip-grams naturally protect from most popular rephrasing techniques:

1. Replacing words by synonyms – skip-grams will just "skip" the replaced word and engine will detect a match of skip-gram constructed from previous and next words.
2. Removal of words – since there are skip-grams with and without removed word the engine will find a match.
3. Addition of words – since the skip-grams applied on detection phase too, there are skip-gram that will be matched to skip-gram from indexed document without this word.
4. Changing the order of words – words in a skip-gram are sorted in alphabetical order, so the order of words within document will not affect the detection.

As can be seen on Figure 5, both methods received as input the same text segment *"Barack Obama has issued an invitation to Israeli Premier Benjamin Netanyahu to visit the White House."*, that is a rephrasing of previously fingerprinted *"Barack Obama invites Netanyahu for White House visit"* (Figure 5). The new method detected three matches between these text segments, while full fingerprint did not detect even a single match.

## 5. PRELIMINARY EVALUATION

The proposed method was evaluated using two different datasets and on three scenarios. We chose to compare our new method with the well-known full fingerprinting method [24] described in Section 3. Full fingerprinting is recognized as one of the most effective methods for detecting plagiarism [24] and for near duplicate detection [47]. In this section we present the evaluation process and results.

### 5.1 Evaluation Measures

In order to evaluate the proposed method when executed with different parameters and compare it with the full fingerprinting method, we used the area under ROC curve measure (AUC). ROC is a graph representation of the tradeoff between the true positive rate (TPR) and the false positive rate (FPR) for different thresholds. TPR and FPR are computed using the following four measurements:

- *True positive (TP)* is the number of documents containing confidential content actually classified as confidential.
- *False positive (FP)* is the number of documents not containing confidential content that are incorrectly classified as confidential.
- *False negative (FN)* is the number of documents containing confidential content that are incorrectly classified as non-confidential.
- *True negative (TN)* is the number of documents not containing confidential content actually classified as non-confidential.

The true positive rate and the false positive rate (i.e., the rate of false alarms) are computed are follows:

$$TPR = \frac{TP}{TP+FN} \qquad FPR = \frac{FP}{TN+FP}$$

To compare the effectiveness of the different settings of the proposed method or between different methods we compare the results of computing the area under ROC curve (AUC) for each of the methods (which is in the range of [0, 1]).

In addition, we measured the space efficiency and performance of the evaluated methods. We define the space

*efficiency* measure as the number of characters in all fingerprinted documents divided by the number of records in the fingerprints DB (in evaluation of the proposed method, only fingerprints of confidential document were considered). Similarly, *performance* is defined as the average number of characters per second that can be processed during the detection phase.

Both methods were implemented in C#[2] using .Net 4.0 framework[3] with in-memory fingerprints DB. We used a PC with Intel Core i5-2500K CPU and 4GB of memory for our experiments.

## 5.2 Datasets

Since we could not find publicly available datasets for evaluating data leakage methods, our evaluation was conducted using two datasets which we adjusted to our needs:

- **Reuter news** - a dataset that was compiled from news articles collected from the Reuters news[4] feed for a duration of two months. The documents of the dataset belong to 17 different categories ranging from art and culture to economics and sports and based on the IPTC news codes[5]. The number of documents in the entire dataset was 6,102. We chose to define the documents of a single sub-topic of the economics topic: "international (foreign) trade" as confidential. The number of "international (foreign) trade" articles was 310.
- **Subset of PAN Plagiarism Corpus 2010** - PAN PC-2010 [35] contains automatically and manually generated plagiarism cases of passages from books obtained from Project Gutenberg[6]. In order to generate a large number of human made plagiarism cases, Potthast M. *et al*. [46] employed an Amazon's Mechanical Turk[7] (a marketplace for work that requires human intelligence). The goal of the participants in the task was to rephrase a given text while preserving the meaning of the original text:

  *"Rewrite the original text found below so that the rewritten version has the same meaning as the original, but with a different wording and phrasing. Imagine a scholar copying a friend's homework just before class, or imagine a plagiarist willing to use the original text without proper citation."*

  We extracted from PAN-PC-10 3,036 human rephrased passages which belong to 30 different, automatically generated clusters. The clustering should have been by topic, but since the diversity of the books obtained from the Gutenberg Project was too large, documents that belong to the same cluster may not share a topic in a way that is obvious to a human onlooker. Although the goal of this dataset is to evaluate plagiarism detection systems, it is the largest dataset with manually created paraphrases that we found which we could use to evaluate the ability of our system to identify rephrased documents.

## 5.3 Evaluated Scenarios

The proposed method was evaluated within the following scenarios:

- **Scenarios 1:** evaluating the detection of **short** (up to 50 words), **rephrased part** (that contained the main idea of each document) of a **previously seen** confidential document embedded within a larger **non-confidential document**. This scenario was evaluated on the "Reuter news" dataset. The dataset for evaluation was created as follows. For the indexing phase, we used all "confidential" documents (total of 310) and 66% of the "non-confidential" documents (total of 3791). For the detection phase, we used the remaining 33% of the "non-confidential" documents (total of 1954). In addition, from the remaining 33% of the "non-confidential" documents, we randomly selected 310 documents and within each file we embedded a rephrased version of a short part of the "confidential" document. Due to the lack of resources for human rephrasing we rephrased the documents artificially by switching every third word with a synonym using Microsoft Word's[8] Synonym finder. This process is illustrated in Figure 6.

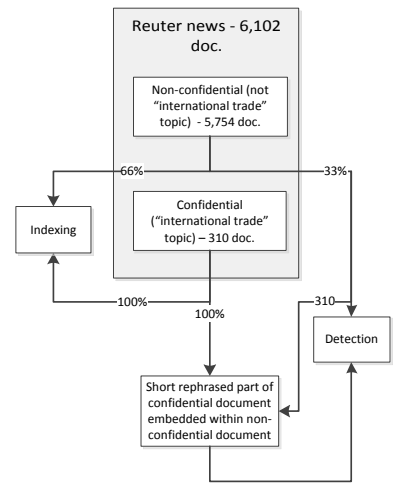

**Figure 6. Scenario 1: creation process**

- **Scenario 2:** evaluating the detection of **short** (up to 50 words) **part** (that contained the main idea of each document) of previously **unseen** confidential document embedded in a larger **non-confidential** document. This scenario was evaluated on the "Reuter news" dataset as well. The dataset for evaluation was created as follows. For the indexing phase, we used 66% of the "confidential" documents (total of 206) and 66% of the "non-confidential" documents (total of 3791). For the detection phase, we used the remaining 33% of the "non-confidential" documents (total of 1954). In addition, from the remaining 33% of the "non-confidential" documents, we randomly selected 104 documents and within each file we embedded a short part of the "confidential" document from the remaining 104 "confidential" documents. This process is illustrated in Figure 7.
- **Scenario 3: evaluating the detection of human made rephrasing**. This scenario was evaluated on the PAN PC-

---

[2] http://msdn.microsoft.com/en-us/vstudio/hh388566
[3] http://msdn.microsoft.com/en-us/library/zw4w595w.aspx
[4] http://www.reuters.com/
[5] http://www.iptc.org/site/NewsCodes/
[6] http://www.gutenberg.org/
[7] http://www.mturk.com
[8] http://office.microsoft.com/en-us/word/

2010 dataset. Cluster number 29, containing 377 text passages, was defined as confidential. The testing set contained corresponding rephrased versions of those passages as confidential documents and rephrased versions of passages in the other 29 clusters (2,659 documents) as non-confidential. In this scenario we attempted to detect a rephrasing of the whole confidential document. Additionally, the confidential cluster contains a diversity of different topics, making it more difficult to distinguish between confidential and non-confidential documents. This process is illustrated in Figure 8.

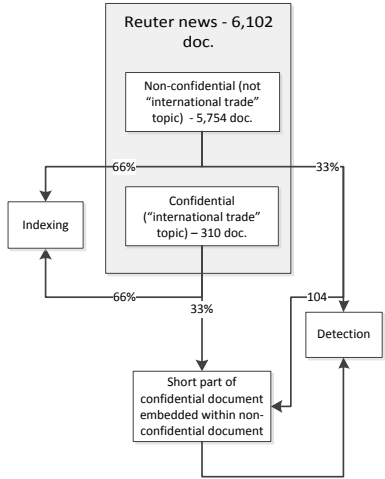

**Figure 7. Scenario 2: creation process**

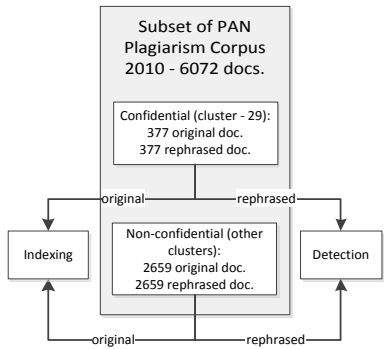

**Figure 8. Scenario 3: creation process**

The reason that we tested scenarios 1 and 2 only on the Reuters news dataset is that the PAN-PC-2010 dataset is mostly based on novels and extracting a short part that describes the main idea of such a type of text is in most cases impossible. In contrast, the Reuters news dataset is based on news articles which are naturally short and it is therefore feasible to find short segments of text which summarize the main idea of the article.

We also evaluated scenario 3 on the Reuters news dataset, as well as full, near, and partial duplicate detection scenarios (described in section 2) on both datasets; however, since both the full fingerprinting and the proposed method achieved perfect accuracy in these scenarios, we chose not to present the results.

## 5.4 Configurable Parameters
Full fingerprinting has the following configurable parameters:

- *Stemming* – applying stemming on words from input document. Can be enabled or disabled.
- *Stopwords removal* – removing stopwords from input document. Can be enabled or disabled.
- $n$ – number of words in $n$-gram.

In addition to the above parameters, the new method has the following parameters:

- *Sorting* – alphabetically sorting the words within $k$-skip-$n$-grams. Can be enabled or disabled.
- $k$ – the number of skips allowed in $k$-skip-$n$-grams.
- $m$ – the minimal number of non-confidential documents containing specific $k$-skip-$n$-gram, so it should be filtered out of the fingerprints database.

## 5.5 Evaluation Results
The results of comparing the two evaluated methods are presented in Table 2. The table depicts for each combination of scenarios (1, 2, and 3) the detection method (full fingerprinting and the proposed method) and the configuration settings (stemming, stopwords removal, $n$, sorting, $k$ and $m$), the following measures: AUC, space efficiency, and performance. It should be noted that for each scenario we performed an evaluation of all the possible combinations of parameters of both methods, totaling in about 3000 experiments for each scenario. However, we only present the most interesting configurations of both methods.

In general, the results show that the proposed fingerprinting method outperformed the full fingerprinting method for almost all configurations, achieving significantly better AUC, even when configured to be as space efficient as full fingerprinting. Despite the fact that better detection accuracy comes at cost of performance, the proposed method still provide acceptable performance to be implemented in real-time environment.

In scenario 1, both methods achieved the best results when stemming and stopwords removal were disabled. The new method achieved a prefect AUC, being even more space efficient and as fast as full fingerprinting. It should be noted that in this simple scenario using $k$-skip-$n$-grams instead of regular $n$-grams has little effect on detection results. The ROC graphs of the full fingerprinting (id=2) and the new method (id=4) are presented in Figure 9.

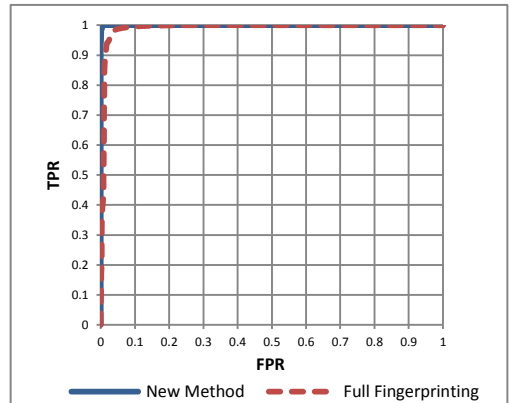

**Figure 9. Scenario 1: ROC**

For scenario 2, both methods achieved the best results when stopwords removal is enabled. Stemming, however, improves the detection accuracy of full fingerprinting and degrades it when applied to the new method. In this scenario, the new method achieved good detection accuracy results, much better than full fingerprinting even when k-skip-n-grams is not used (id=11). As opposed to the previous scenario, here better results can be achieved when the number of skips in *k*-skip-*n*-grams is set to a higher value. The ROC for the full fingerprinting (id=8) and the new method (id=14) are presented in Figure 10.

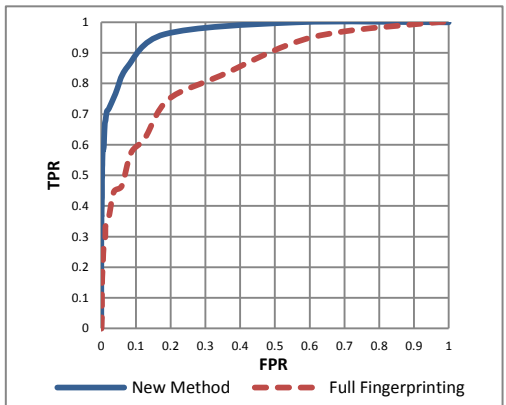

**Figure 10. Scenario 2: ROC**

For scenario 3, both methods achieved the best results when stemming is enabled. Stopwords removal improves the detection accuracy of full fingerprinting and degrades it when applied to the new method. In this scenario the new method achieves very good detection accuracy. Unfortunately, it is achieved at the expense of space efficiency. The ROC graphs for full fingerprinting (id=15) and the new method (id=21) are presented in Figure 11.

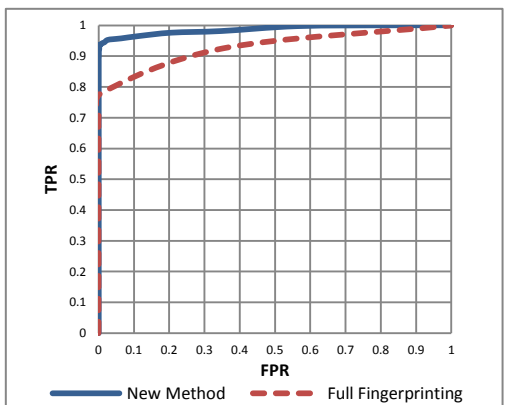

**Figure 11. Scenario 3: ROC**

It can be seen that the parameters n - (number of words in the n-gram), and *k* - (number of skips allowed) that yield the best AUC for each scenario are different, therefore they should be adjusted for each specific domain. Figures 12, 13, 14 show the effect of *n* and *k* on AUC for each scenario. Even though the charts differ significantly, there is one interesting consistent phenomenon that occurs in all scenarios; the effect of *n* on AUC decreases with the increase of k. Therefore, when the properties of a domain are not known in advance, an organization should set *k* relatively high in order to compensate a probable wrong choice of *n*. In general, it seems that setting *n* to 3 and setting relatively high *k* is the best default configuration that was appropriate for most of the tested domains.

In addition, we noticed that for the proposed method, the stemming and the stopwords removal have very little effect on AUC (much lower than in full fingerprinting). This is a positive outcome because we can save processing time by disabling stemming; stopwords can be removed to reduce the fingerprint database size without a significant impact on AUC. Another possible advantage of this insensitivity to the text preprocessing is that the proposed method can be applied for multi-language data leakage detection. However, this subject should be further studied.

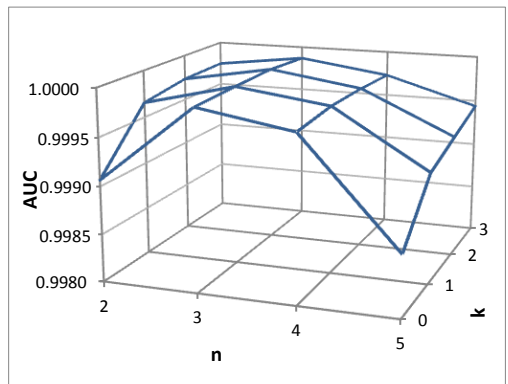

**Figure 12. Scenario 1: AUC**

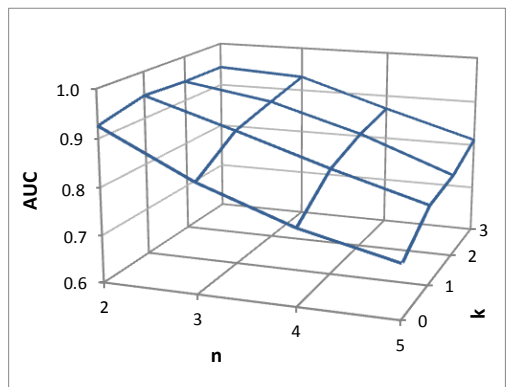

**Figure 13. Scenario 2: AUC**

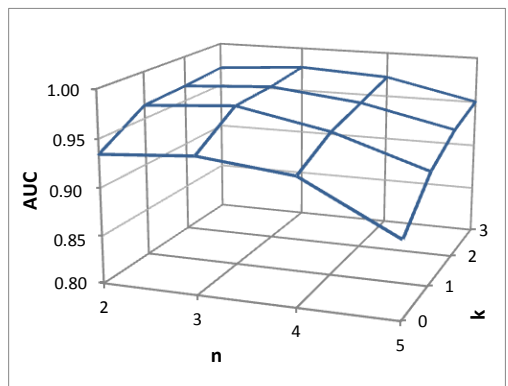

**Figure 14. Scenario 3: AUC**

Table 2. Evaluation results.

| | | Experiment ID | Configuration | | | | | | Results | | |
|---|---|---|---|---|---|---|---|---|---|---|---|
| | | | Stemming | Stopwords Removal | n | Sorting | k | m | AUC | Space efficiency (char./record) | Performance (char./second) |
| Scenario 1 | Full fingerprinting | 1 | disabled | disabled | 3 | | | | 0.988075 | 6.236 | 3181255 |
| | | 2 | disabled | disabled | 4 | | | | 0.990827 | 6.181 | 3124342 |
| | | 3 | disabled | disabled | 5 | | | | 0.985997 | 6.182 | 2978276 |
| | New Method | 4 | disabled | disabled | 3 | disabled | 0 | 1 | 0.999879 | 8.150 | 3140811 |
| | | 5 | disabled | disabled | 3 | disabled | 1 | 1 | 0.999896 | 2.608 | 2344703 |
| | | 6 | disabled | disabled | 3 | disabled | 2 | 1 | 0.999906 | 1.294 | 1458476 |
| | | 7 | disabled | disabled | 3 | disabled | 3 | 1 | 0.999868 | 0.780 | 1115541 |
| Scenario 2 | Full fingerprinting | 8 | enabled | enabled | 2 | | | | 0.842483 | 12.036 | 2156117 |
| | | 9 | enabled | enabled | 3 | | | | 0.831987 | 11.820 | 2125230 |
| | | 10 | enabled | enabled | 4 | | | | 0.765432 | 11.868 | 2100985 |
| | New Method | 11 | disabled | enabled | 2 | enabled | 0 | 1 | 0.925723 | 16.546 | 2004016 |
| | | 12 | disabled | enabled | 2 | enabled | 1 | 1 | 0.948979 | 8.445 | 1732869 |
| | | 13 | disabled | enabled | 2 | enabled | 5 | 1 | 0.954353 | 3.221 | 1053185 |
| | | 14 | disabled | enabled | 2 | enabled | 9 | 1 | 0.963276 | 2.158 | 756366 |
| Scenario 3 | Full fingerprinting | 15 | enabled | enabled | 2 | | | | 0.923151 | 16.709 | 883021 |
| | | 16 | enabled | enabled | 3 | | | | 0.846854 | 16.795 | 883930 |
| | | 17 | enabled | enabled | 4 | | | | 0.775764 | 17.070 | 874032 |
| | New Method | 18 | enabled | disabled | 3 | enabled | 0 | 1 | 0.941407 | 7.299 | 724957 |
| | | 19 | enabled | disabled | 3 | enabled | 1 | 1 | 0.970740 | 2.582 | 544331 |
| | | 20 | enabled | disabled | 3 | enabled | 4 | 1 | 0.982527 | 0.619 | 205159 |
| | | 21 | enabled | disabled | 3 | enabled | 6 | 1 | 0.986497 | 0.366 | 110734 |

## 6. CONCLUSION

In this paper we present a modified version of full fingerprinting [24] that uses *sorted k-skip-n-grams*. We show that the proposed fingerprinting method is more robust to the rephrasing of confidential content and can filter non-relevant (non-confidential) parts of a document. We evaluate the proposed method and illustrate its superiority over the full fingerprinting approach, while maintaining acceptable execution time.

*Sorted k-skip-n-grams* achieve higher detection accuracy, compared to the *n*-gram-based fingerprinting, especially in the unintentional leakage scenario, and in the "simple" (unsophisticated) malicious attacker scenario. A determined attacker will be able to evade the *k*-skip-*n*-grams, for example by encoding words; however, this is a limitation of any existing data leakage detection system or method. Note that if a malicious attacker encrypts a file with an attempt to leak, a common practice is to block any outgoing content that cannot be analyzed by the data leakage detection system [40]. In addition, in order to counteract this type of attack, different anomaly detection techniques can be integrated in order to provide a comprehensive protection.

In addition, most of DLD systems contain a data discovery module [40]. The goal of this module is to discover un-indexed confidential content on organization's servers and employees' computers. The proposed method may significantly improve the accuracy of such module, since it is much better in detection of previously unseen confidential content.

The proposed method achieves better detection accuracy when configured to be as space efficient as full fingerprinting. However, one should consider sacrificing space efficiency in order to further improve results. We believe that some organizations might prefer to invest money to satisfy the storage requirements for an improved data leakage detection system rather than absorb significant financial and reputational losses caused by un-detected and hence un-prevented data leakage accidents.

The space requirements become an acute limitation of the real life implementation of the proposed method when the detection is performed at the endpoints. In modern DLD systems, an agent installed on an employee's computer requires a detection of data leakage attempts even when the end-point is not connected to the organization's network. In such a scenario, synchronizing large fingerprints database between the DLD server and agent can become unfeasible.

In light of these drawbacks, our future research will focus on improving the space efficiency of the proposed method, preferably conserving its high detection accuracy and acceptable performance.

In addition, we plan to extend our experiments and evaluate the proposed method using a new dataset that should better resemble real life data leak accidents. Also, the effect of noise in input documents tagging on detection accuracy should be studied. Finally, we plan to explicitly evaluate the handling of standard content (disclaimers, standard form, common phrases, etc.) within confidential documents.